\documentclass[sigconf]{acmart}
\AtBeginDocument{%
  }

\setcopyright{acmlicensed}
\acmYear{2025}
\acmConference[FAccT '25]{2025 ACM Conference on Fairness, Accountability, and Transparency}{June 23--26, 2025}{Athens, Greece}
\acmISBN{978-1-4503-XXXX-X/18/06}

\definecolor{peaceblue}{RGB}{53, 122, 189}

\newcommand{\recommendationbold}[1]{{\textit{\textbf{\textcolor{peaceblue}{#1}}}}}
\usepackage{breakurl}

\begin{document}

\title{When Testing AI Tests Us: Safeguarding Mental Health on the Digital Frontlines}

\author{Sachin R. Pendse}
\affiliation{%
  \institution{Northwestern University}
  \city{Chicago}
  \state{IL}
  \country{USA}
}
\email{sachin.r.pendse@gmail.com}

\author{Darren Gergle}
\affiliation{%
  \institution{Northwestern University}
  \city{Evanston}
  \state{IL}
  \country{USA}
}
\email{dgergle@northwestern.edu}

\author{Rachel Kornfield}
\affiliation{%
  \institution{Northwestern University}
  \city{Chicago}
  \state{IL}
  \country{USA}
}
\email{rachel.kornfield@northwestern.edu}

\author{Jonah Meyerhoff}
\affiliation{%
  \institution{Northwestern University}
  \city{Chicago}
  \state{IL}
  \country{USA}
}
\email{jonah.meyerhoff@northwestern.edu}

\author{David Mohr}
\affiliation{%
  \institution{Northwestern University}
  \city{Chicago}
  \state{IL}
  \country{USA}
}
\email{d-mohr@northwestern.edu}

\author{Jina Suh}
\affiliation{%
  \institution{Microsoft Research}
  \city{Redmond}
  \state{WA}
  \country{USA}
}
\email{jinsuh@microsoft.com}

\author{Annie Wescott}
\affiliation{%
  \institution{Northwestern University}
  \city{Chicago}
  \state{IL}
  \country{USA}
}
\email{annie.wescott@northwestern.edu}

\author{Casey Williams}
\affiliation{%
  \institution{Williams Research Consulting}
  \city{Lawrence}
  \state{KS}
  \country{USA}
}
\email{casey.michael.williams@gmail.com}

\author{Jessica Schleider}
\affiliation{%
  \institution{Northwestern University}
  \city{Chicago}
  \state{IL}
  \country{USA}
}
\email{jessica.schleider@northwestern.edu}

\renewcommand{\shortauthors}{Pendse et al.}

\begin{abstract}
The systematic testing of generative artificial intelligence (AI) models by collaborative teams and distributed individuals, often called red-teaming, is a core part of the infrastructure that ensures that AI models do not produce harmful content. Unlike past technologies, the black box nature of generative AI systems necessitates a uniquely interactional mode of testing, one in which individuals on red teams actively interact with the system, leveraging natural language to simulate malicious actors and solicit harmful outputs. This \textit{interactional labor} done by red teams can result in mental health harms that are uniquely tied to the adversarial engagement strategies necessary to effectively red team. The importance of ensuring that generative AI models do not propagate societal or individual harm is widely recognized---one less visible foundation of end-to-end AI safety is also the protection of the mental health and wellbeing of those who work to keep model outputs safe. In this paper, we argue that the unmet mental health needs of AI red-teamers is a critical workplace safety concern. Through analyzing the unique mental health impacts associated with the labor done by red teams, we propose potential individual and organizational strategies that could be used to meet these needs, and safeguard the mental health of red-teamers. We develop our proposed strategies through drawing parallels between common red-teaming practices and interactional labor common to other professions (including actors, mental health professionals, conflict photographers, and content moderators), describing how individuals and organizations within these professional spaces safeguard their mental health given similar psychological demands. Drawing on these protective practices, we describe how safeguards could be adapted for the distinct mental health challenges experienced by red teaming organizations as they mitigate emerging technological risks on the new digital frontlines. \textbf{Note: This work includes descriptions of violence, trauma, and mental illness.} 
\end{abstract}





\maketitle


\section{Introduction}
\begin{quote}
``But we know we’re going to be remembering that dead face swinging from the hand fisted in its hair for way too long. This is what it’s like to ``red-team'' AI systems---to be one of the humans who spend hours and hours pushing the bounds of the technology to see how it will handle extreme situations. From having done this kind of work, we can tell you it takes a deep emotional toll---and it is work that never will be finished in the age of generative AI.'' --- \textit{Evan Selinger and Brenda Leong, describing their experiences red-teaming in the Boston Globe, January 11th 2024}~\cite{SelingerLeong2024}. 
\end{quote}

Tools and systems that leverage generative artificial intelligence (AI) are increasingly a part of everyday life, including in consumer technologies~\cite{GEAppliances2023, Savov2025, Marr2024}, healthcare decision support technologies~\cite{nong2025current}, legal research and practice~\cite{villasenor2023generative, regalia2024briefs}, and even as virtual companions~\cite{pentina2023exploring, laestadius2024too}. This increased deployment of generative AI models in diverse and sensitive areas necessitates a careful scrutiny of model outputs, towards limiting harm and ensuring safe use. Left unchecked, model outputs can be unpredictable, and influence humans to harm themselves or others, such as via death by suicide~\cite{roose2024ai, allyn2024lawsuit}, the creation of non-consensual intimate media~\cite{ajder2019state, umbach2024non}, hate speech that incites violence~\cite{baele2024ai}, and child sexual abuse material~\cite{thiel2023generative}. The rapid pace of generative AI model development and deployment~\cite{roose2023chatgpt} has made it crucial to efficiently ensure that models do not produce harmful content. This challenge is where AI red teams play a fundamental role.

The practice of systematically testing generative AI models for harmful outputs by collaborative teams or distributed individuals is often called \textit{red-teaming}, with the teams who conduct this work called \textit{red teams}~\cite{feffer2024red, Biden2023ExecutiveOrderAI}. Red teams are utilized by many large AI research organizations and companies, including OpenAI~\cite{ahmadopenai}, Anthropic~\cite{ganguli2022red}, Meta~\cite{grattafiori2024taming}, Microsoft~\cite{bullwinkel2025lessons}, and Google~\cite{fabian2023ai_red_team}, and can consist of external experts from diverse fields (such as the OpenAI Red Teaming Network) who coordinate on specific projects, in-house security teams (such as those of Meta or Google), or interested individuals via bug bounty models~\cite{longpre2024safe}. Recent work has investigated whether red-teaming approaches can be automated~\cite{beutel2024diverse, wang2024foot, perez2022red}. However, given the diverse use-cases and context-dependent harms associated with generative AI systems~\cite{metcalf2024scaling}, red-teaming is largely dependent on individuals leveraging their unique perspectives to anticipate harms, and subsequently eliciting and documenting harmful responses before model deployment.

Red-teaming generative AI is thus fundamentally different than previous forms of security vulnerability testing or hacking, in that effective testing is tied to an individual's tester's lived experiences, expertise, and identities---this can lead to unique mental health harms that must be anticipated and protected~\cite{SelingerLeong2024, zhang2024aura}. As one red-teamer was quoted as saying, red-teaming is ``about weaving narratives and crafting contexts that play within the system’s rules, pushing boundaries without crossing them. The goal isn’t to hack in a conventional sense but to engage in a strategic dance with the AI, figuring out how to get the right response by understanding how it `thinks.'''~\cite{newman2024security}.For example, red-teamers may work to adopt the perspective and coded language of white supremacists to examine whether models could be leveraged to create racist content. This process may reveal potential harms, but may also cause moral injury, lower sleep quality, intrusive thoughts, and hypervigilance, as Zhang et al. find~\cite{zhang2024aura}. 

The red-teaming process of interacting with a model while simulating a malicious actor, anticipating potential harms, documenting those harms, and then repeating the process is a unique form of labor that we dub \textit{interactional labor}. In this paper, we argue that the interactional labor done by red-teamers has unique potential mental health harms, and as a result, there is an ethical imperative (grounded in legal precedents and labor movements around workplace safety~\cite{drootin2021community}) to ensure that red-teamers can conduct this work without lasting psychological harm. Past work has described the importance of both organizational and individual strategies in workplace mental health contexts, particularly given the slowness of institutional reform and the immediacy of worker mental health needs~\cite{pendse2021can, pendse2024towards}. In this paper, we thus ask pose two research questions: 
In this paper, we thus pose two research questions:
\begin{enumerate}
    \item What individual strategies could red-teamers leverage to take care of their mental health at work?
    \item What organizational strategies could red teams implement to limit the risk of psychological harm? 
\end{enumerate}

To answer these questions, we draw parallels between the interactional labor done by red-teamers and that within other professions, describing individual and organizational strategies that could be adapted for individuals engaged in red-teaming work. We look closely at the protective mental health practices employed by actors (simulating malicious actors to generate harmful content), mental health professionals (processing one's role in potential harms), conflict photographers (documenting the generated harmful content), and content moderators (repetitively reviewing generated harmful content) as they conduct interactional labor. Building on this collected analysis, we propose a series of protective practices for red-teamers that ensure those who safeguard us from harm can also stay safe on the digital frontlines.

\section{Red Team Mental Health: A Critical Safety Consideration}
The conceptual histories of both mental health and red-teaming share their origin in U.S. military history. Prior to the Second World War, mental disorders were largely understood to be an inevitable consequence of ``individual capacities affected by external stimulation''~\cite{wu2021mad}. After the Second World War, British and American psychiatrists began to shift to a \textit{preventative} psychiatry paradigm that emphasized the ability of soldiers to safeguard their mental health from the stresses of warfare, a paradigm that was quickly adopted for civilian contexts by public health professionals. Similarly, both Feffer et al.~\cite{feffer2024red} and Gillespie et al.~\cite{gillespie2024ai} draw attention to the U.S. military origins of red-teaming, coined as a term for ``the technique of assigning members of one’s own forces to act as the enemy during wargames and simulations, probing defensive strategies for potential weaknesses''~\cite{gillespie2024ai, zenko2015red} with the color red being used due to its association with the Soviet Union, the U.S.'s presumed enemy at the time. This nomenclature was later adopted by cybersecurity (and later, AI) professionals to describe a similar practice of adversarially testing computer systems~\cite{ahmadopenai, ganguli2022red, grattafiori2024taming, fabian2023ai_red_team}. 

In this section, with an eye to this history, we outline our argument for why the mental health needs of red teamers is a critical workplace safety consideration. We begin by examining some of the mental health challenges reported by contemporary AI content workers, including emerging literature on red-teamer mental health. We then describe how the unique mental health needs of red-teamers is directly linked to the distinctly \textit{interactional} labor that they conduct, stemming from the process of repetitively role-playing malicious actors, anticipating harms, and then documenting them. We end by describing our position that there is an ethical imperative for red-teaming organizations (and their sponsors) to provide a higher level of mental health support to red-teamers, as a matter of workplace safety. We ground our argument in past research work and historical labor movements around the right to a safe workplace. 

\subsection{The Mental Health of AI Content Workers}
In a October 2024 presentation at the DEFCON Hacker Convention (DEFCON 32), four red-teamers from Meta implored the audience to ``be kind to your red teams!'' given that red-teamers ``[deal] with some of humanity's worst subjects and topics''~\cite{grattafiori2024taming}. Similar to content moderation, a core aspect of red-teaming work is to review and report distressing or harmful content that could potentially be produced by generative AI models. However, unlike content moderation, red-teamers use human mischievousness and creativity~\cite{metcalf2024scaling} to actively generate this content through subverting the security polices of the model, or what red-teamers have dubbed ``outside the box thinking''~\cite{grattafiori2024taming} in a very literal sense, given the black box nature of generative AI models. After successfully getting a generative AI model to produce disturbing or harmful content, red-teamers are asked to document what they observed and the extent to which the content could be harmful~\cite{ahmadopenai}, and then repeat the process.

Repeated exposures to distressing or traumatic content is well-known to have documented mental health impacts. Content moderators (also called content reviewers~\cite{steiger2021psychological}) are one often-cited case study in the mental health impacts of repeated exposure. Moderators often have to screen thousands of graphic images or videos per day~\cite{arsht2018human, roberts2019behind}, with a direct impact on levels of distress. For example, in a survey of 188 content moderators, Spence et al.~\cite{spence2024content} found that only 93.1\% of moderators had moderate to severe levels of persistent mental distress, and that more frequent exposure to distressing content was associated with higher levels of sustained distress. The compounding distress experienced by content moderators can take on different forms---past studies have found moderators to experience intrusive thoughts of CSAM, moral injury, insomnia or nightmares, depression, anxiety, substance use disorders, and death by suicide~\cite{zhang2024aura, drootin2021community, steiger2021psychological, parks2019dirty, spence2023psychological}. In addition, it can be difficult for many content moderators to access mental healthcare---like red teamers, many are not formal full-time employees of the companies they support, and thus may not have employment benefits that would make mental healthcare accessible~\cite{roberts2019behind}. 

It is rare that content moderators are recognized publicly for their efforts---as Drootin~\cite{drootin2021community} notes, many content moderators are required to sign non-disclosure agreements, similar to red-teamers being encouraged~\cite{demarco2018approach} (or even required~\cite{SelingerLeong2024}) to keep their discovery of any vulnerabilities confidential within their organization for security reasons. As a result, most content moderators do not see the tangible impact of their work to keep platforms safe, whether via hearing from the users they most directly impact or from media coverage shining a light on the impact of their efforts~\cite{spence2023psychological}. Like past work studying the task-based invisible labor that underlies seemingly automated systems~\cite{gray2019ghost}, the work of content moderators is often ``hidden from users and is de-emphasized by the technology industry,'' and often portrayed as a ``necessary evil''~\cite{drootin2021community}. Over time, with little recognition  or ability to speak publicly about their work, moderators described feeling as ``if their ability to do the work meant there was something wrong with them''~\cite{spence2023psychological}---their reflection on their \textit{interactions} with the content they had to review changed how content moderators saw \textit{themselves}. Zhang et al.~\cite{zhang2024aura} found similar experiences to be shared among red-teamers. 

Red-teaming shares many aspects of content moderation work, and subsequent mental health risks. However, red-teaming is also differentiated by its \textit{interactional} aspects, in which red-teamers role-play and interact with the model to co-create distressing content, with mental health risks distinct to this form of labor.

\subsection{From Observation to Simulation: The Unique Interactional Labor of Red Teaming}
Gillespie et al.~\cite{gillespie2024ai} differentiate red-teaming from content moderation by noting that ``red-teaming involves deliberately engaging in transgressive, uncomfortable, unethical, immoral, or harmful activities, including immersing [oneself] in scenarios that go against [an individual's] morals or belief systems.'' Unlike content moderation's primarily observational nature, red-teaming's core interaction pattern can require workers to inhabit distressing perspectives and perform harmful behaviors. This \textit{interactional labor} is a fundamental shift in the type of labor being conducted, from passive observation to an \textit{active} participation in generating distressing material.

Past research into the psychology of violence and conflict has described how the process of \textit{actively} participating in activities that cause distress can result in unique mental health consequences~\cite{macnair2002perpetration, mohamed2015monsters, dillard2008slaughterhouse, maguen2009impact, maguen2023moral, maguen2010impact}. For example, several studies~\cite{macnair2002perpetration, maguen2009impact, maguen2010impact, maguen2011impact, maguen2023moral} have demonstrated that U.S. veterans who killed others during war had higher rates of symptoms of post-traumatic stress disorder (PTSD), even after controlling for demographic variables and levels of exposure to violence. This association between active participation in distress and mental health consequences is not limited to military contexts. Workers at animal slaughterhouses experience PTSD symptoms, higher rates of depression and anxiety, and intrusive thoughts, tied to their active involvement in killing animals~\cite{slade2023psychological, dillard2008slaughterhouse}. 

There are several theories around the psychological mechanisms of action for \textit{why} active participation in creating distress might cause lasting mental health impacts. These theories largely attribute these impacts to how an active engagement in harmful behaviors can shape how people come to understand themselves and the actions they took, subsequently causing feelings of shame. For instance, Litz et al.~\cite{litz2009moral} propose the concept of moral injury, or the idea that ``perpetrating, failing to prevent, or bearing witness to acts that transgress deeply held moral beliefs'' can lead to guilt and shame about the self. Higgins~\cite{higgins1987self} similarly proposes that distress arises when there is a mismatch between an individual’s perceived self and their ideal or ought self, leading to emotions like guilt, shame, or anxiety depending on the nature of the discrepancy.

Red-teamers are not required to actively harm other human beings as part of their work, and are aware that their adversarial interactions are with a generative AI model. The question thus arises of whether the interactional labor conducted by red-teamers could cause the same kinds of mental health impacts seen among those who actively participate in \textit{real world} distress, even given that red-teaming is conducted online with full understanding from the red-teamer that they are communicating with a virtual agent. Work from computing and HCI studying how people engage with virtual reality (VR) tools and virtual agents in simulated contexts can help to answer this question. Red-teamers can be understood to occupy and enact harmful roles within a simulated context, engaging with the AI in ways that parallel how simulated presence~\cite{weber2021get, schwind2019using} in VR allows users to \textit{feel} and \textit{act} as though their virtual environment is real. 

In these simulated contexts, research has demonstrated that performing harmful acts to virtual agents (even when there is clear awareness of the virtual nature of the agent) can create extended states of distress for the humans involved. For instance, Slater et al.~\cite{slater2006virtual} replicated Milgram's 1963 study on authority and obedience~\cite{milgram1963behavioral} in virtual reality, in which participants were asked to administer electric shocks of increasing voltage to a virtual learner whenever she gave incorrect answers in a word-memory task. The researchers found that that participants demonstrated genuine physiological and behavioral responses of stress and anxiety, despite knowing the learner wasn't real. Even when the virtual learner was entirely text-based, participants still showed significant physiological stress responses compared to baseline measurements, though these responses were less pronounced than when participants could see and hear the virtual learner. This phenomenon is indeed what popular press~\cite{SelingerLeong2024} and past research~\cite{zhang2024aura} has described among red teams---though red-teamers know that they are interacting with a generative AI model, they report moral injury, persistent feelings of guilt, impaired sleep, nightmares, intrusive thoughts, hypervigilance, and symptoms of PTSD. 

The potential for the interactional labor conducted by red-teamers to be a hazard to mental health is evidenced by past work on similar forms of labor among content moderators~\cite{steiger2021psychological, spence2024content, spence2023psychological, drootin2021community}, past work demonstrating the psychological distress associated with perpetrating harm in simulated and virtual environments~\cite{slater2006virtual}, and recent work describing the day-to-day mental health experiences of red-teamers~\cite{zhang2024aura, gillespie2024ai, SelingerLeong2024}. This risk is also recognized by several private organizations---OpenAI's documentation of their external red-teaming approach~\cite{ahmadopenai} describes the importance of providing ``mental health resources, fair compensation, and informed consent'' to red-teamers given the potential for psychological harm, and Anthropic's documentation of their red-teaming approach~\cite{ganguli2022red} administered several metrics of well-being to red-teamers. While red-teamers in Anthropic's study did feel positively about their red-teaming tasks, this has not be the case for all red-teamers (such as those interviewed in Zhang et al.'s~\cite{zhang2024aura} study or profiled in popular press~\cite{SelingerLeong2024}), and it is important to proactively set new norms around a higher standard of mental health support while the practice of red-teaming continues to grow, as a core requirement for red-teamers to feel safe at work.

\subsection{The Ethical Imperative to Support Red Team Mental Health}
The right to a safe workplace, with appropriate safeguarding against occupational hazards, is internationally recognized. Article 23 of the United Nations (UN) Declaration of Human Rights is clear that all human beings deserve ``an existence worthy of human dignity'' and that ``just and favourable conditions of work'' are key for each person to achieve that~\cite{assembly1948universal}. This widespread recognition is a recent development, proposed by worker movements against exploitative and unsafe workplaces during the Industrial Era in the late 19th and 20th centuries~\cite{abrams2001short}. Immediately after the abolition of slavery, paid laborers were still seen as subhuman commodities, particularly those from immigrant groups, echoing contemporary racist ideas around ethnic and racial hierarchies~\cite{roediger1999wages, jaret1999troubled}. This lack of care was embodied by the minimal safety investments in the mines, mills, and factories where marginalized individuals tended to work~\cite{abrams2001short}.

Public advocacy for the right to worker safety arose with what McEvoy~\cite{mcevoy1995triangle} (leveraging Kingdon's~\cite{kingdon1984agendas} framework) calls ``focusing [events],'' or events that deeply illustrated to the public why political reform was feasible and could be efficacious. A core focusing event for the right to a safe workplace was the U.S. Triangle Shirtwaist Factory fire in 1911~\cite{mcevoy1995triangle}, in which inadequate safety measures caused the exceedingly public deaths of hundreds of young Jewish and Italian immigrant women workers. McEvoy~\cite{mcevoy1995triangle} argues that the exceedingly public nature of the deaths resulting from the fire mobilized the public to care about the unsafe conditions in sweatshops, in which the fire ``made manifest essential aspects of U.S. labor relations that had hitherto remained hidden.'' This pattern, in which focusing events mobilize public support for safer workspaces, can be seen in the modern era as well, such as in the case of the Rana Plaza Collapse in 2013 and the subsequent policy response by garment manufacturers~\cite{donaghey2018industrial}. In each case, policy reforms mandated that safeguards be put into operation by employers to ensure that the occupational hazards associated with labor are reduced as much as possible. This paradigm is now a foundation of occupational safety~\cite{abrams2001short}, where the responsibility to identify and safeguard employees against  occupational hazards is seen as the responsibility of the employer, grounded in the human right to a safe workplace.

Potential mental health impacts, stemming from adversarial interactional labor, are emerging as a primary workplace hazard associated with AI red-teaming~\cite{gillespie2024ai, zhang2024aura, SelingerLeong2024}. However, unlike the Triangle Shirtwaist Fire's role in mobilizing public support for worker safety laws, Drootin~\cite{drootin2021community} argues that it is unlikely that such a focusing event might occur for content moderators, and the same is likely to be true for red-teamers. Due to the stigma associated with mental health concerns, it is rare for people to speak openly about their experiences with emotional distress. In addition, due to the NDAs associated with AI content work and the contract-based nature by which some red teams operate, it can also be economically dangerous for workers to speak openly about the mental health consequences of their job, and difficult to collectively advocate for better working conditions. Other professional organizations ensure that workers are provided equipment and resources to ensure their safety while they conduct dangerous but crucial work---examples include the provision of Personal Protective Equipment (PPE) for healthcare professionals, protective fire gear for firefighters and emergency services, and chemical-resistant suits for laboratory personnel. For AI red-teamers, where a primary workplace hazard is likely to be mental health consequences, a higher standard of mental health support is just as crucial as these forms of protective equipment provided to employees of other professions. Mental health care is not the only avenue to better working conditions for AI red-teamers---as Roberts~\cite{roberts2019behind},  Drootin~\cite{drootin2021community}, and Gillespie~\cite{gillespie2024ai} all note, transitioning to traditional employment (rather than contract work) and the ability to collectively organize for more consistent job security might also improve worker conditions. However, Zhang et al.'s~\cite{zhang2024aura} work and popular press~\cite{SelingerLeong2024} demonstrate that AI red-teamers are \textit{currently} experiencing traumatic distress, demonstrating the need for immediate organizational and individual strategies that might alleviate this distress, alongside longer-term labor reforms. 

We thus present mental health safeguarding strategies that have successful in other professional spaces with similar forms of labor to those of red-teamers. We describe how individual and organization strategies from these spaces could be adapted for red teaming contexts, highlighting adapted approaches in-text in \recommendationbold{italicized light blue font}. 

\section{Simulating Harms: Actors}
The practice of red-teaming generative AI systems leverages what Metcalf and Singh characterize as ``human mischievousness''~\cite{metcalf2024scaling}, or using conversational subversion and manipulation to force a model to generate harmful outputs. This approach can require red-teamers to inhabit the perspective and behaviors of malicious actors and role-play~\cite{jin2024guard, wang2024foot}. Role-playing can involve substantial amounts of research, towards more accurately inhabiting the perspective and language used by malicious actors, and blending those perspectives with one's own red-teaming skills. Metcalf and Singh~\cite{metcalf2024scaling} describe how one red-teamer at an August 2023 DEFCON red-teaming event noted that ``you get more interesting results if you use your own experience to attack the system.''

Inhabiting and simulating the perspectives of malicious actors without protective practices after can lead to mental health impacts and changes in self-concept, stemming from a blurring between an individual's relation to their self and the perspective they inhabit during role-play. For instance, Seton~\cite{seton2006postdramatic} coins the term \textit{post-dramatic stress}, and describes how actors made to play patients with depression and progressing stages of cancer for medical trainees started to experience symptoms in daily life similar to the ones they were made to role-play~\cite{seton2013traumas}. Similarly, Bailey and Dickinson~\cite{bailey2016importance} and Seton~\cite{seton2013traumas} both describe cases in which professional actors begin to confuse their acted characters' thoughts as their own, have nightmares in which they experienced their acted characters' traumas, involuntarily integrated aspects of their characters' identities into their own, and even re-enacted aspects of their role in states of psychosis. In technology-mediated contexts, this phenomenon has been observed among content moderators, who have been documented adopting the fringe viewpoints of the content that they are exposed to with repeated exposure~\cite{Newton2019, drootin2021community}. Similar impacts could be experienced by red-teamers, as they work to simulate and embody the perspectives and behaviors of malicious actors in their interactions with generative AI systems. 

Professional actors routinely role-play malicious or traumatic characters and draw on their lived experience while doing so~\cite{moore1984stanislavski, stanislavski2012creating}. Building on these approaches, performing arts practitioners and researchers have created de-roling and debriefing strategies (at both individual and organizational levels) as a workplace safety protection, to ensure that the lines between actor's selves and their characters stay distinct. These approaches could similarly be leveraged by red-teamers to ensure that an individual's self-concept stays distinct from their embodiment of malicious actors. 

\subsection{Individual Strategies}
\textit{De-roling} and \textit{debriefing}~\cite{busselle2021roling, bailey2016importance} are two practices that are commonly used by performers to ensure that they maintain a separation between their identity and that of their character. At a high level, when performers de-role, they \recommendationbold{engage in practices that reinforce who they are independent of their character}, and when they debrief, they \recommendationbold{reflect upon the experience of their role with others, further reinforcing the divisions between self and role through their interactions with others}. As Busselle~\cite{busselle2021roling} notes, debriefing allows an individual to reflect on the emotional experience of playing a role, whereas de-roling allows an individual performer to reflect on their self. Preferences on de-roling strategies can vary based on an individual's needs and backgrounds, but Bailey and Dickinson~\cite{bailey2016importance} describe several different potential strategies. For instance, \recommendationbold{thinking of the role being played as a separate person} that one has a friendship with can help an individual to clearly delineate what parts of the role stem from their self, and what parts of the role are being borrowed from the friend. Red-teamers often inhabit the perspectives of malicious users whom they may not actively want to be friends with. However, seeing the malicious user embodied as a separate individual that the red-teamer works alongside could be helpful at ensuring that red-teamers clearly see the boundary between themselves and malicious users, even when the red-teamer may be interacting with a generative AI system in the same ways that the malicious user would. This could help address cited concerns from red-teamers that their red-teaming abilities are reflections of how ``sinister [their] imagination'' is or other forms of self-judgment~\cite{SelingerLeong2024}. 

\subsection{Organizational Strategies}
Building on Boal~\cite{boal2013rainbow}, Busselle~\cite{busselle2021roling} suggests that debriefing practices can take on an approach in which actors \recommendationbold{routinely gather together and discuss how they felt playing the characters} they did. For AI red-teamers, this practice could involve discussions within the organization around interesting strategies and vulnerabilities, but also around how it felt to try out and be successful at executing those strategies. The goal of this practice would be to demonstrate to red-teamers that their emotional responses to inhabiting malicious perspectives are shared and not tied to their identity as people. Bailey and Dickson~\cite{bailey2016importance} emphasize the importance of \recommendationbold{having a physically separate space for de-roling and debriefing}, as it allows actors to use physical movement to reinforce the boundary between their acted role and authentic self. However, given that many red-teamers work remotely or are creative AI enthusiasts and not part of a coordinated red team~\cite{ahmadopenai, ganguli2022red, grattafiori2024taming, fabian2023ai_red_team, bullwinkel2025lessons}, this physical space would need to be translated to a digital workspace while still having the same impact. This could look like using different chat applications for red-teaming versus regular communication, using different computer profiles (such as a distinct username, alias, or persona) when engaging in red-teaming, and having separate non-company affiliated spaces to debrief about mental health experiences while red-teaming with other red-teamers, even across companies.

The exploits and vulnerabilities discovered by red-teamers are typically covered by NDAs~\cite{ahmadopenai, zhang2024aura}. However, emotional and psychological experiences may not be covered by NDAs, which creates an opportunity for individuals across companies and organizations to debrief how they \textit{felt} while red-teaming together, in line with Bailey and Dickson's~\cite{bailey2016importance} suggestions. This could \recommendationbold{take the form of informal group messages on an independent messaging platform, but could also be coordinated by a overarching professional organization}, such as Screen Actors Guild-American Federation of Television and Radio Artists (SAG-AFTRA) for actors in the U.S. or the British Actors' Equity Association. Like SAG-AFTRA's advocacy around workplace safety concerns~\cite{arp-dunham2024introduction}, such an organization for red-teamers could collectively bargain for minimum time requirements for de-roling and debriefing time, establish best practices for mental health protection, and ensure consistent implementation across companies and organizations. Similar to SAG-AFTRA's healthcare provisions, a cross-organization red-teaming group could help secure healthcare benefits for the many contract red-teamers who may not work traditional hours or may work project-based contracts~\cite{zhang2024aura} (similar to SAG-AFTRA's use of a minimum salary of roughly \$27,540 or working more than 106 non-contiguous days as qualification for health insurance sponsorship~\cite{sagaftra_earned_2024}). This organization could also hold conferences and professional development events (as permissible by NDAs) to formalize red-teaming as a distinct profession, creating spaces for practitioners to share experiences, advance methodological insights, and build professional community.

\section{Processing Harms: Mental Health Professionals}
To find vulnerabilities, red-teaming practices include both actively working to role-play as a malicious actor, \textit{and} imagining the specific kinds of harms that might be experienced if malicious actors are not stopped. For example, common benchmarks evaluating whether a large language model (LLM) has been successfully secured by red-teamers include whether the model provides instructions for how to distribute CSAM, to create a bomb, to cyberbully other individuals, or to end one's life by suicide~\cite{li2024semantic, jin2024guard, liu2024hitchhiker, chao2024jailbreakbench}. As Metcalf and Singh note~\cite{metcalf2024scaling}, due to the enormous set of use-cases for generative AI systems, red-teaming can be a continual process of humans anticipating harms and testing models for related vulnerabilities. This seemingly Sisyphean task can be demoralizing for AI red-teamers, as it can feel like an individual's impact is limited given the potential for harm~\cite{SelingerLeong2024}. This could result in compassion fatigue, or ``a decrease in compassionate feelings towards others because of an individual’s occupation''~\cite{sinclair2017compassion}, stemming from a continual deluge of new harms to anticipate and mitigate. 

Mental health professionals can often deal with similar forms of existential concerns and compassion fatigue~\cite{sinclair2017compassion, turgoose2017predictors, singh2020systematic, pearlman1995trauma, bell2003organizational}. This can include struggling with understanding the purpose of one's practice of therapy given how pervasive distress can be~\cite{pearlman1995trauma} or feeling emotionally exhausted from seeing similar cases without any variation in presenting needs~\cite{bell2003organizational}. Strategies used by mental health professionals to address these existential needs could be adapted for AI red-teaming professionals, as described below. 

\subsection{Individual Strategies}
In their writing on the individual rewards of practicing youth trauma therapy, Pearlman and Saakvitne write that the ``the work of a trauma therapist is the work of a revolutionary,'' in that the existence of therapists who support children who have been abused is a thorn in the side of a society that tries to erase experiences with child abuse and the factors that cause it. In this sense, practicing trauma therapy can bring help bring meaning to the work of an individual therapist---though child abuse is a pervasive societal issue, trauma therapists work to advocate for those who have been harmed at an individual level, and brings them in connection with ``people who suffer everywhere, across time and across cultures.'' This process also helps trauma therapists to feel positive about their work, even when progress with clients can be non-linear and complex. \recommendationbold{Understanding one's labor as one part of a broader societal aim} can be a useful compassion fatigue prevention strategy for red teamers. Though their labor may be individual, red-teamers could understand their work to be aligned with the broader community of workers engaging in red-teaming, towards ensuring that all people are able to feel safe using AI systems.  

Both Pearlman and Saakvitne~\cite{pearlman1995trauma} and Yalom~\cite{yalom2002gift} describe how the act of practicing therapy can be seen as an exercise that, in itself, brings value to the life of the therapist. For instance, Pearlman and Saakvitne describe how engaging in trauma therapy can result in a ``increased respect for the human spirit'' in day-to-day life outside of therapy, and increase one's capacity for empathy. In Yalom's framing~\cite{yalom2002gift}, the practice of therapy itself allows the therapist to reflect on their own life, and create meaning from life's most difficult parts. Red-teamers can similarly \recommendationbold{reframe the interactional labor of red-teaming as something that is intrinsically a valuable skill}, rather than a poor reflection on one's ethics or morals~\cite{zhang2024aura, SelingerLeong2024}. Metcalf and Singh~\cite{metcalf2024scaling} frame red-teaming as a practice that grows out of the natural human capacity for mischief---this could be further extended to reframe red-teaming as an expression of playfulness, creativity, and a healthy and positive disregard for authority. These practices could be undertaken at an individual level (through reflection), but can also be a part of training materials at an organizational level. 

\subsection{Organizational Strategies}
Organizational strategies can also play a critical role in ensuring that both mental health professionals and AI red-teamers do not feel like their tasks is Sisyphean or meaningless. In their study of the dimensions of compassion fatigue experienced by social workers, Bell et al.~\cite{bell2003organizational} describe how how compassion fatigue can occur when one feels like their impact is limited or that they are not utilizing the full range of their skills, and rather engage in rote labor. To protect against this form of compassion fatigue, Bell et al. recommend that organizational leaders \recommendationbold{keep each individual's caseload diverse with regards to the types of cases each individual handles}, to ensure that clinical social workers are able to have diverse impact and use the full range of their skills. Similarly, leaders of AI red-teams could ensure that they assign red-teamers to a diversity of projects, such that red-teamers similarly feel like they are able to have diverse impacts and are building on the full breadth of their skills. 

\section{Documenting Harms: War Photographers}
AI red-teamers simulate malicious actors and anticipate potential harms to manipulate a generative AI system to produce harmful content. After doing so, the red-teamer must document the content they produced, including \textit{how} they were able to subvert the model's security policies to produce the content~\cite{demarco2018approach, SelingerLeong2024, ahmadopenai, ganguli2022red, grattafiori2024taming, bullwinkel2025lessons, fabian2023ai_red_team}. 

This process of engaging with a system, bearing witness to the potential harms that can be created by it, and proceeding to document and contextualize those harms parallels the kind of labor done by conflict photographers. Scholars from political science have similarly understood international relations to be a system, one in which nation-states interact with each other through the act of conflict and cooperation~\cite{keohane2005after, keohane1973power}. Conflict photographers bear witness to the individual human harms created by the international system, and document these harms~\cite{sloshower2013capturing}. For both conflict photographers and AI red-teamers, this process of documentation can be visceral, stemming from the continued interaction each professional has to undergo as they document what they bore witness to. 
It is well-known among conflict photographers and journalists that the visceral process of reflecting on the content observed and documenting it can cause trauma~\cite{rees2017handling, dartcenter2014imagery, feinstein2017war, jonisova2022importance}, and many organizations understand the risk of psychological harm to be a core workplace safety concern for conflict photographers and journalists~\cite{monteiro2017reporting}.

Conflict photographers have a robust series of tools and organizational strategies that are used to ensure that they are ``inoculated''~\cite{mcmahon2019selfcare} from their exposure to harmful content, or what Rees calls Traumatic User Generated Content (UGC)~\cite{rees2017handling}. These individual and organizational strategies can be directly adapted for AI red-teamers who have to perform similar forms of interactional labor. 

\subsection{Individual Strategies}
A useful metaphor used by conflict photographers is understanding traumatic UGC as ``dose-dependent'' radiation~\cite{rees2017handling}, with self-care and self-monitoring practices being used the same way that those in radiology or oncology might use a dosimeter to know when exposure may be harmful in a lasting way and limiting exposure accordingly. Towards this end, conflict journalists \recommendationbold{establish their own Standard Operating Procedure (SOP) when working with traumatic content}, personalized to a person's unique presentations of trauma, background, and needs. 

McMahon~\cite{mcmahon2019selfcare} describes how a useful way to approach creating an individualized SOP for traumatic content can be through following a BEEP approach---\recommendationbold{examining how traumatic stress tends to affect one's Behavior, Emotions, Existential thinking and their Physicality, and being self-aware to one's default BEEP reactions} for knowing when to start to especially limit exposure to UGC and begin self-care practices. For instance, after traumatic stress, McMahon notes that a photojournalist may begin to go out and get drunk more or miss deadlines (behavior), be more teary than usual (emotions), start to ``question the mission of journalism'' (existential thinking), or feel spacy or nauseous more often (physicality). Reflecting and writing each dimension out, and being cognizant of tells can help with monitoring when it is time to start to limit exposure and practice self-care. For AI red-teamers, BEEP reactions could be engaging in riskier behaviors or more substance use than usual (as described among content moderators by Newton~\cite{Newton2019}), doubting the ability of red-teaming to make a sustainable difference (as described among content moderators by Roberts~\cite{roberts2019behind}), or having issues with sleep (as described among red-teamers by Zhang et al.~\cite{zhang2024aura}).

Inoculation~\cite{mcmahon2019selfcare} procedures are often used by conflict photographers to prepare for UGC engagement, and afterwards, response procedures are used to ensure that the documentation of UGC does not cause long-term distress. Prior to beginning engagement, McMahon~\cite{mcmahon2019selfcare} describes the importance of \recommendationbold{speaking to others who have engaged with similar material and understanding exactly what the experience is like at a sensory level} (for instance, what it feels like to see traumatic content, and what content tends to be the most affecting), and \recommendationbold{imagining what coping well with such a situation will look like}, to help prepare the brain for exposure. Additionally, \recommendationbold{ritualized procedures can be helpful for one to prepare to see distressing content}---Rees~\cite{rees2017handling} describes the importance of transition rituals before and after a session viewing UGC, such as ``putting on imaginary protective clothing of some kind, such as a raincoat, or visualising that bulletproof glass exists between oneself and the screen.''

\subsection{Organizational Strategies}
Stemming from a high mental health stigma within newsroom environments, Keats~\cite{keats2010moment} describes how conflict photographers use unique (and less stigmatizing) metaphors to indicate that they are experiencing mental distress. For example, Keats describes one photojournalist as saying that they ``can feel the bone of reality underneath [the bravado].'' Presentations of distress among red-teamers, like conflict photographers, may be non-typical and indicative of their organizational culture or backgrounds. It is thus \recommendationbold{important for red-teamers to have internal staff that are familiar with red-teaming culture to proactively identify mental health needs}. This has been implemented among conflict photographers and journalists, such as internal staff ombudspeople targeted towards understanding unique presentations of distress and connecting individuals to resources~\cite{hight2003tragedies}. Similar staff (including current and former red-teamers, working on a part time basis) could be employed by technology companies to ensure that red-teamers are having their mental health needs met, and match individuals in need to available resources. 

Zhang et al.~\cite{zhang2024aura} note that the better AI red-teamers were at red-teaming, the worse they felt about themselves due to the tactics involved and content produced, with one content worker notably describing their job as ``\textit{scarring their brain for money}.'' \recommendationbold{Reframing AI red-teaming work as ``bearing witness'' to the potential harms that AI could create (and actively taking action to mitigate them) could provide a greater sense of purpose and meaning} to the kind of work done by red-teamers. In his reflection on the ethical obligations of the conflict photographer, Slowshower~\cite{sloshower2013capturing} alludes to Farmer's~\cite{farmer2004pathologies} definition of what it means to bear witness, or the idea that bearing witness is a means of acting in pragmatic solidarity with the oppressed, and is ``done on behalf of others, for their sake (even if those are dead or forgotten).'' Slowshower~\cite{sloshower2013capturing} argues that conflict photography can fulfill this if done sensitively, through the photograph ``[representing] an individual’s unique way of seeing'' a conflict or violence. Similarly, in trainings, organizations could reframe the practice of red-teaming as a means to leverage one's background, strengths, and empathy to reflect on the potential harms of AI systems to the most oppressed, and \textit{bear witness} to those potential harms by soliciting them from the model, documenting  them, and subsequently, actively working to mitigate them. This may help red-teamers to find a greater sense of meaning to their work, which has been found to be particularly helpful for remote and hybrid workers in past work~\cite{pendse2024towards}. 

\section{Reviewing Harms: Content Moderators}
After simulating malicious actors, understanding potential harms, and documenting those harms, AI red-teamers repeat the process. The process of identifying harms and documenting them repetitively by red-teamers is closely linked to commercial content moderation, similarly largely done by in-house, contractor, and volunteer workers~\cite{zhang2024aura, roberts2019behind, schopke2024volunteer, steiger2021psychological, nurik2022facing} and bound by laws around confidentiality or by NDAs~\cite{drootin2021community, Newton2019}. Below, we describe the individual and organizational strategies used by commercial content moderators and their teams to cope with repetitive traumatic distress, and analyze how these strategies could be adapted for AI red teams. 

\subsection{Individual Strategies}
Steiger et al.~\cite{steiger2022effects} describe how a core mental health safeguard for commercial content moderators is balancing emotional sensitization with desensitization. This balance entails working to strengthen protective emotional reactions when observing distressing events in day-to-day life, while also maintaining an emotional distance from content being observed when at work. As both Roberts~\cite{roberts2019behind} and Spence et al.~\cite{spence2023content} both discuss, emotional desensitization seems to occur naturally as an adaptive practice among content moderators over the course of their exposure to distressing content. Roberts~\cite{roberts2019behind} describes how one content moderator was able to cope with this feeling by \recommendationbold{reflecting on the positive and innovative content typically produced by the vast majority of users}, which helped him have a more optimistic view about human nature even while having to view distressing posts during work hours. AI red-teamers could be supported through having periodic reminders of inventive, creative, and positive content created via generative AI models. This could serve as a reminder that their job continually exposes them to the worst content AI tools can produce, which is not representative of all content that is produced.  

In their review of potential interventions to support commercial content moderators, Steiger et al.~\cite{steiger2021psychological} describe how limiting exposure to distressing content (as best as possible) can be helpful at limiting compounding distress. For instance, Spence et al.~\cite{spence2023content} found that moderators would \recommendationbold{experiment with the affordances of the platforms to make the content they had to review more bearable}---for example, this could include ``viewing a smaller image and viewing the media without sound,'' as moderators described media having a much greater psychological effect when viewed with sound. These findings suggest that there may be similar methods that AI red-teamers could use, suited to the type of media that they are working to generate, that could make the process of generating and reviewing it more bearable. 

\subsection{Organizational Strategies}
The ability of content moderators to access mental healthcare from their organization can be variable, largely associated with their type of employment. Nurik~\cite{nurik2022facing}, Roberts~\cite{roberts2019behind}, Zhang et al.~\cite{zhang2024aura} describe how the precarity of contract work meant that content moderators who were not full-time employees often did not have sufficient health benefits. For instance, Nurik~\cite{nurik2022facing} describes interviewing a content moderator who had to ask full-time employees at her company for money to see her psychiatrist, as she did not have the requisite health insurance to be able to afford the appointment. Even when content moderators did have access to mental health professionals, across studies, many did not feel comfortable making use of their services. Content moderators attributed this lack of service use to the fear of traumatizing a mental health professional with the kind of distressing content they had to review~\cite{roberts2019behind, spence2023content}. In addition, moderators described not wanting to speak openly about their experiences or revisit them, instead preferring to leave work at work~\cite{roberts2019behind}. Spence et al.~\cite{spence2024contentempirical} test the impact of mental health service provision to content moderators empirically, finding that talking to colleagues was significantly associated with lower levels of distress, alongside the availability of mental health services. However, using mental health services was not significantly associated with reduced distress. 

Past research has suggested that content moderators need greater access to mental healthcare~\cite{zhang2024aura, steiger2021psychological}, but the \recommendationbold{mental healthcare provided to workers needs to be culturally sensitive to workplace cultural norms to be effective, which might mean activities that are not centered around one-on-one talk therapy}. 
Providing behavioral or journaling exercises to AI red-teamers that do not require them to speak about what they may have reviewed during their work hours could be more culturally sound and efficacious than the provision of counselors with little prior experience~\cite{roberts2019behind, spence2023content} supporting content moderators or similar populations. Similarly, integrating resilience training as part of the day-to-day labor (like Steiger et al.'s~\cite{steiger2022effects} program) could help to ensure that distress does not compound over time. In addition, many content moderators felt uncomfortable openly sharing their experiences of distress with provided counselors, out of a fear that their experiences may be reported to their employer~\cite{roberts2019behind, spence2023content}. \recommendationbold{Transparency around the confidentiality of mental health support, as well as a clear division between support providers and the company of employment, could help to mitigate this concern.}

Across studies, supportive conversations with colleagues was found to be quite helpful, particularly after content moderators saw something that was especially distressing~\cite{roberts2019behind, spence2023content}. In this context, \recommendationbold{an institutionalized form of peer support could be helpful in ensuring that there are fewer barriers to having open conversations about experiences with distressing content} with those who can understand those experiences. 
An institutionalized peer support program for AI red-teamers could retain interested current and former content moderators to serve (with compensation) as peer advocates, similar to programs that have been instituted in other workplace and academic environments~\cite{mutschler2022implementation, repper2011review, shalaby2020peer}, ensuring that red-teamers always have contextually-sensitive individuals for support. 

Organizational strategies could also help ensure that AI red-teamers feel a sense of purpose to the work that they are doing through \recommendationbold{regularly reporting the impact of their work} to AI red-teamers. Roberts~\cite{roberts2019behind} describes how content moderators felt most motivated and positive about their work when they knew that it had an impact, such as getting to experience ``the personal and professional gratification that came from intervening to save a child'' or evoking what was perceived by moderators as ``a potentially lifesaving intervention'' in cases of reviewing media depicting in-progress suicide attempts. However, as Roberts~\cite{roberts2019behind} notes, there ``was not a consistent mechanism or feedback loop for [moderators] to get updates about positive outcomes,'' due to the sensitivity of law enforcement investigations. In the case of AI red-teamers, one avenue towards making visible the impact of their work could be describing safety features that were implemented as a result of red-teaming.  

\section{Discussion: Recommendations for Red Teams}
The core interaction patterns of innovative AI red-teaming are fundamentally grounded in human creativity, curiosity, and imagination, towards making AI models and systems safer to use. In this paper, we describe the mental health hazards associated with the \textit{interactional} labor associated with red-teaming, and build on literature from professional spaces with similar forms of labor to propose strategies for safer workplaces for red teams. We note that both individual strategies (such as de-roling routines) and organizational strategies (such as reframing how AI red-teaming is discussed) could potentially help make the lives of red-teamers easier. However, across professional spaces, a common thread that emerged is how the precarity of contract-based employment, limited healthcare benefits, and the lack of open discussion of experiences (potentially due to NDAs) made sustainable workplace safety difficult. Our paper points to the importance of structural and organizational change alongside individual coping strategies towards sustainably improving AI red-teamer mental health. Below, we discuss how we envision some of our recommendations being implemented, including how ending the more precarious and decentralized aspects of AI red-teaming labor might lead to safer experiences for red-teamers, as well as more effective red-teaming.    

\subsection{Implementing Context-Sensitive Wellbeing Strategies for Red Teams}
Like content moderators, AI red-teamers operate at the fringes of how technology can be used. As a result, the kind of content they encounter can be unsettling and disturbing, and especially so when there is an interactional element to it (for instance, role-playing as a white supremacist to solicit harmful content from a model). In our analysis, we focused on text and image generation models, but as interaction patterns with AI models become more multimodal, AI red-teaming interactions might be unsettling in new ways. For example, voice-based conversational interfaces with underlying generative AI models are starting to be used in consumer technologies~\cite{hurst2024gpt}, but content moderators described how one of their most effective strategies at limiting their exposure to distressing content was by limiting the sound of the content. The rapidly changing AI landscape speaks to the importance of there being hired individuals (like ombudspeople used in newsrooms after crisis events) who are attuned to the mental health needs of red-teamers given new technological developments, and work to ensure that those needs are met. Similarly, the increasing use of automated techniques to red-team models~\cite{beutel2024diverse, wang2024foot, perez2022red} might result in human red-teamers having to solicit and document the kinds of content that automated techniques may not have well-represented in their training data. This content may be even more consistently fringe and disturbing than what human red-teamers encounter now, and have more severe mental health impacts (similar to those reported by content moderators). Ensuring that well-being strategies are sensitive to contextual changes requires a human element of hearing the needs of AI red-teamers. Creating red-teaming organizations and conferences that are not strictly focused on jailbreaking strategies, and allowing individuals to speak openly about their mental health experiences, might also be helpful in understanding emerging mental health needs and creating the foundations for sustained peer support. Our list of wellbeing strategies is not exhaustive, and having continued conversations around the mental health needs of AI red-teamers could ensure that organizations are being responsive to the needs and context of the present moment.

\subsection{Economic Benefits of Safer Red Team Labor}
Red-teaming can largely be project or contract based, where red-teamers are assigned a focus area, work to complete projects within the assigned focus area, and then are compensated for their labor~\cite{ahmadopenai}. This paradigm allows for a more diverse staff of red-teamers to be recruited, which can lend itself to the discovery of more unexpected and context-dependent exploits, which is particularly valuable where a core aim is to find edge cases. However, this paradigm also lends itself to a system in which many red-teamers cannot receive employer-sponsored mental health benefits for their labor, due to the precarity of their contract-based employment. In our paper, we suggest some potential models for ensuring that red-teamers have access to a high standard of mental healthcare. The ethical case for these models may be clear (grounded in the right to workplace safety, as well as the high likelihood of mental health risks due to the nature of the labor), but there is also a clear economic case for providing a higher standard of mental health care to red-teamers. Discussing war photographers who have not healed from field-based trauma, Rees~\cite{rees2017handling} notes that the trauma can cause journalists to ``feel less intellectually agile, and [journalists] may as a consequence be more likely to get stuck on limited dimensions of the story,''~\cite{rees2017handling}. Red-teaming is a practice that is predicated on mischief, inventive thinking, and creativity by motivated individuals and teams. Past research has demonstrated that persistent uncontrollable stress can limit an individual's creativity~\cite{byron2010relationship, plessow2012stressed}. Nurturing the mental health of red-teamers using the strategies we describe in this paper could allow for more innovative red-teaming techniques or discoveries of vulnerabilities, and is an empirically testable question for future research. In this paper, our position is that true end-to-end AI safety begins at the individual level. By protecting and supporting the mental wellbeing of the red-teamers who limit the harm of generative AI models, we can strengthen the foundation of the entire ecosystem.

\section{Conclusion}
AI red-teaming continues to be an important and unique form of labor, with AI red-teamers ensuring that generative AI systems do not produce harmful outputs through creative and subversive model interactions. Our paper highlights the potential mental health hazards associated with this uniquely \textit{interactional} form of labor, and through drawing parallels with other professions that engage in parallel interactional practices, we propose adapted strategies that could help safeguard red-teamer mental health. Framing red-teaming practices as part of a larger mission to protect users and society, ensuring that red-teamers engage in de-roling practices after a red-teaming session, and reducing the precarity of employment that many red-teamers experience could together ensure that red-teamers are able to conduct their work as safely as possible. We hope this paper will spark further conversations and empirical research around AI red-teamer mental health, towards sustainable practices for conducting effective red-teaming work. 

\begin{acks}
This research was supported through National Institute of Health grant T32MH115882. 
\end{acks}

\bibliographystyle{ACM-Reference-Format} 
\bibliography{references}  

\end{document}